\documentclass[conference]{IEEEtran} 

\usepackage{amsmath}
\usepackage{graphicx}
\usepackage{booktabs}
\usepackage{fancyhdr}
\usepackage{amsfonts}
\usepackage{url}
\usepackage{algorithm}
\usepackage{algorithmic}

\fancypagestyle{firstpage}{
  \fancyhf{}
  \fancyhead[L]{Accepted for publication at the 10th ACM International Conference on Intelligent Systems, Metaheuristics \& Swarm Intelligence (ISMSI 2026), April 24-26, Cebu City, Phillipines.}
}

\title{An Information-Theoretic Framework for Comparing Voice and Text Explainability}
\author{
\IEEEauthorblockN{Mona Rajhans}
\IEEEauthorblockA{
    Senior Manager, Software Engineering\\
    Palo Alto Networks\\
    mrajhans@paloaltonetworks.com
}
\and
\IEEEauthorblockN{Vishal Khawarey}
\IEEEauthorblockA{
    Staff Software Engineer\\
    Quicken Inc\\
    vishal.sanfran@gmail.com
}
}

\begin{document}
\maketitle
\thispagestyle{firstpage}
\pagestyle{plain}

\begin{abstract}
Explainable Artificial Intelligence (XAI) aims to make machine learning models transparent and trustworthy, yet most current approaches communicate explanations visually or through text. 
This paper introduces an information-theoretic framework for analyzing how explanation modality—specifically, voice versus text—affects user comprehension and trust calibration in AI systems. 
The proposed model treats explanation delivery as a communication channel between model and user, characterized by metrics for \textit{information retention}, \textit{comprehension efficiency} ($CE$), and \textit{trust calibration error} ($TCE$). 
A simulation framework implemented in Python was developed to evaluate these metrics using synthetic SHAP-based feature attributions across multiple modality–style configurations (brief, detailed, and analogy-based). 
Results demonstrate that text explanations achieve higher comprehension efficiency, while voice explanations yield improved trust calibration, with analogy-based delivery achieving the best overall trade-off. 
This framework provides a reproducible foundation for designing and benchmarking multimodal explainability systems and can be extended to empirical studies using real SHAP or LIME outputs on open datasets such as the UCI Credit Approval or Kaggle Financial Transactions datasets.
\end{abstract}

\begin{IEEEkeywords}
Explainable Artificial Intelligence (XAI), Information Theory, Voice AI, Text-to-Speech, Multimodal Explainability, SHAP, LIME, Trust Calibration, Cognitive Modeling, Human–AI Interaction
\end{IEEEkeywords}

\section{Introduction}
\label{sec:introduction}

Explainable Artificial Intelligence (XAI) seeks to make the internal reasoning of machine‐learning systems transparent to human users \cite{doshi2017towards,gunning2019xai}. 
While existing research has produced a wide range of visual and textual explanation methods—such as salience maps, feature attributions, and natural‐language rationales—most of these approaches assume that explanations are \emph{read}, not \emph{heard}.
Yet human communication is inherently multimodal, and voice has long been recognized as an expressive and accessible channel for conveying complex information.

Recent surveys highlight this gap explicitly. 
Akman and Schuller~\cite{akman2024audioxai} observe that, despite rapid progress in visual and textual XAI, \emph{audio explanations remain underexplored} even though they may be more intuitive and expressive for end users. 
Their review builds on the earlier vision of Schuller et al.~\cite{schuller2021sonification}, who proposed ``sonification in multimodal and user-friendly XAI'' and argued that audio explanations could benefit not only audio‐based models (e.g., speech or music analysis) but also non-audio domains where sound serves as an additional or alternative communication channel. 
Together, these works motivate a new line of inquiry: how might the \emph{modality} of explanation—text versus voice—affect the way humans interpret, trust, and act upon AI outputs?

In high-stakes settings such as financial advising or genetic counseling, the form of explanation can shape critical user decisions.
A spoken explanation may foster greater engagement and trust \cite{nass2005wired}, while a textual explanation may support deeper analytical reasoning.
From a human-centered perspective, transparency must also align with user perception and cognitive needs \cite{eiband2018bringing}.
Despite these intuitions, there is little quantitative understanding of how modality interacts with explanation fidelity and cognitive load.

To address this gap, we present a mathematical framework for evaluating modality effects in explainable AI.
We model the transformation of faithful feature attributions (e.g., SHAP \cite{lundberg2017shap} or LIME \cite{ribeiro2016should} values) into human-interpretable messages as an information-transmission process and define metrics for \textit{information retention}, \textit{comprehension efficiency}, and \textit{trust calibration}.
By simulating how these quantities vary across voice and text modalities and different explanatory styles (brief, detailed, analogy-based), we can quantify trade-offs between comprehension and calibrated trust without requiring large user studies.

The contributions of this paper are threefold:
\begin{enumerate}
    \item We introduce an information-theoretic model of modality in XAI, treating explanation delivery as a communication channel between model and user.
    \item We develop quantitative metrics—Comprehension Efficiency ($CE$) and Trust Calibration Error ($TCE$)—to analyze how modality and style influence explanation quality.
    \item We provide simulation results and design insights that align with the emerging field of \emph{audio explainability for non-audio models}, showing how voice and text complement each other in shaping trust and understanding.
\end{enumerate}

Overall, this work contributes to the theoretical foundations of multimodal and human-centered XAI by demonstrating that explainability is not only a matter of \emph{what} is explained, but also of \emph{how} it is communicated.

\section{Methods}
\label{sec:methods}

\subsection{Overview}
We formalize the process of delivering explainable AI (XAI) content through different modalities as an information-transmission problem.
An AI model produces an attribution vector $A$ (e.g., SHAP \cite{lundberg2017shap} or LIME \cite{ribeiro2016should} values) describing how input features contribute to its prediction.
An explanation encoder $E$ transforms $A$ into a human-readable message conditioned on a modality $M\!\in\!\{\text{text},\text{voice}\}$ and an explanation style $S\!\in\!\{\text{brief},\text{detailed},\text{analogy}\}$.
A user with cognitive capacity $C$ receives this message and forms an internal mental representation $U$:
\begin{equation}
    U = f(E(A, M, S), C).
\end{equation}
Our goal is to measure how modality and style affect the fidelity, comprehension, and trust of this communication channel.

\begin{figure}[ht]
    \centering
    \includegraphics[width=0.95\linewidth]{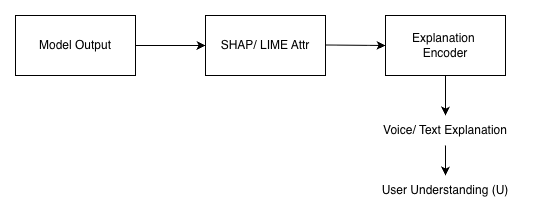}
    \caption{
    \textbf{Conceptual architecture of the proposed modality-aware explainability framework.}
    Model outputs are first processed through SHAP or LIME to obtain attribution vectors ($A$), which are encoded into modality-specific explanations via the explanation encoder $E(A,M,S)$. 
    The resulting message---delivered as text or syntjhesized voice---forms a communication channel with the user, whose understanding ($U$) is analyzed using information-theoretic metrics ($I_M$, $CE$, $TCE$, and $\Phi$).
    }
    \label{fig:pipeline}
\end{figure}

\subsection{Information-Retention Model}
We define the information retention between the model’s true attributions $A$ and the user’s perceived explanation $U$ as the normalized mutual information:
\begin{equation}
    I_M = \frac{H(A) - H(A|U)}{H(A)} = \frac{I(A;U)}{H(A)},
\end{equation}
where $H(\cdot)$ is Shannon entropy \cite{shannon1948mathematical} and $I(A;U)$ quantifies how much information about the original explanation is preserved in the user’s understanding.
A perfect, lossless textual explanation would yield $I_M = 1$; any distortion or omission in the voice rendering would reduce $I_M$.

\subsection{Cognitive-Load Function}
Each modality and style impose a cognitive load $L(M,S)$ on the user:
\begin{equation}
    L(M,S) = \alpha\,D(M,S) + \beta\,H(E(A,M,S)),
\end{equation}
where
\begin{itemize}
    \item $D(M,S)$ is the duration of the explanation (e.g., seconds of audio or number of words),
    \item $H(E(A,M,S))$ approximates the syntactic or conceptual entropy of the message, and
    \item $\alpha,\beta>0$ are modality-dependent scaling constants that model attention fatigue and linguistic complexity.
\end{itemize}
Higher load reduces comprehension efficiency.

\subsection{Comprehension Efficiency}
The Comprehension Efficiency (CE) of an explanation channel is defined as the ratio of retained information to cognitive load:
\begin{equation}
    CE(M,S) = \frac{I_M}{L(M,S)}.
\end{equation}
If $CE_{\text{text}} > CE_{\text{voice}}$, text conveys more usable information per unit of cognitive effort; if the reverse holds, sonification is superior.

\subsection{Trust Calibration}
User trust $T$ should ideally align with the objective correctness of the model’s output $Q$ (e.g., the probability of a correct decision).
We define the Trust Calibration Error (TCE) as
\begin{equation}
    \mathrm{TCE} = \mathbb{E}\big[\,|T - Q|\,\big],
\end{equation}
averaged across all users and tasks.
A smaller TCE indicates better-calibrated trust.
Empirically, prior findings and auditory-perception theory suggest that voice explanations may yield higher $T$ but not necessarily higher $Q$, producing a modest over-trust effect.

\subsection{Overall Evaluation Function}
For a balanced assessment of explanation quality, we integrate comprehension efficiency and trust calibration into a composite score:
\begin{equation}
    \Phi(M,S) = \lambda_1\,CE(M,S) - \lambda_2\,\mathrm{TCE}(M,S),
\end{equation}
where $\lambda_1,\lambda_2$ weight the relative importance of understanding versus calibration.
Maximizing $\Phi$ identifies the modality–-style pair that provides the best trade-off between comprehension and trustworthy engagement.

\subsection{Simulation Protocol}
To validate this model without human participants, we simulate $A$ as synthetic SHAP vectors (e.g., $n = 5$--$10$ features) drawn from realistic distributions for finance and genetics tasks.
We vary explanation entropy and duration to represent different styles and compute the resulting $I_M$, $L(M,S)$, $CE(M,S)$, and $\mathrm{TCE}(M,S)$ under assumed cognitive-capacity parameters $C$.
Plots of $CE$ versus $L$ or $\Phi$ versus $S$ reveal predicted modality effects—voice increasing trust (lower TCE) but text yielding higher CE.

\subsection{Implementation and Evaluation Environment}
The simulation framework was implemented in Python~3.10 using standard open-source libraries including NumPy, pandas, and Matplotlib. 
Synthetic SHAP-based attribution vectors were generated to emulate realistic feature-importance distributions with tunable entropy and duration parameters for each modality--style configuration (Text--Brief, Text--Detailed, Voice--Analogy, etc.). 
Each configuration consisted of 30 simulated samples per domain (finance and genetics), producing 360 total data points. 

All computations were executed on a consumer-grade Intel~i7 CPU with 16~GB RAM, requiring less than three seconds per simulation run. 
Visualization scripts automatically generated comparative plots for Comprehension Efficiency ($CE$), Trust Calibration Error ($TCE$), and the composite score $\Phi(M,S)$. 
This environment ensures full reproducibility and transparency of results while maintaining analytical control over modality parameters. 

Although this study employs synthetic data, the framework is directly compatible with real attributions from models trained on public datasets such as the UCI Credit Approval or Kaggle Financial Transactions datasets. 
By replacing simulated inputs with empirical SHAP or LIME outputs, the proposed pipeline can support empirical evaluation of modality-aware explainability systems in future work.

\section{Results and Discussion}
\label{sec:results}

We simulated SHAP-based feature attributions for synthetic finance and genetics tasks to compare the explanatory effectiveness of \textit{voice} versus \textit{text} modalities across three styles of delivery (Brief, Detailed, Analogy). 
Metrics were computed according to the theoretical framework in Section~\ref{sec:methods}, yielding measures of information retention ($I_M$), cognitive load ($L$), comprehension efficiency ($CE$), and trust calibration error ($TCE$).
Each modality--style combination was evaluated over 360 generated samples (30 per condition per domain), and results were averaged across domains.

\subsection{Trade-off between Comprehension and Trust}
\begin{figure}[ht]
    \centering
    \includegraphics[width=0.9\linewidth]{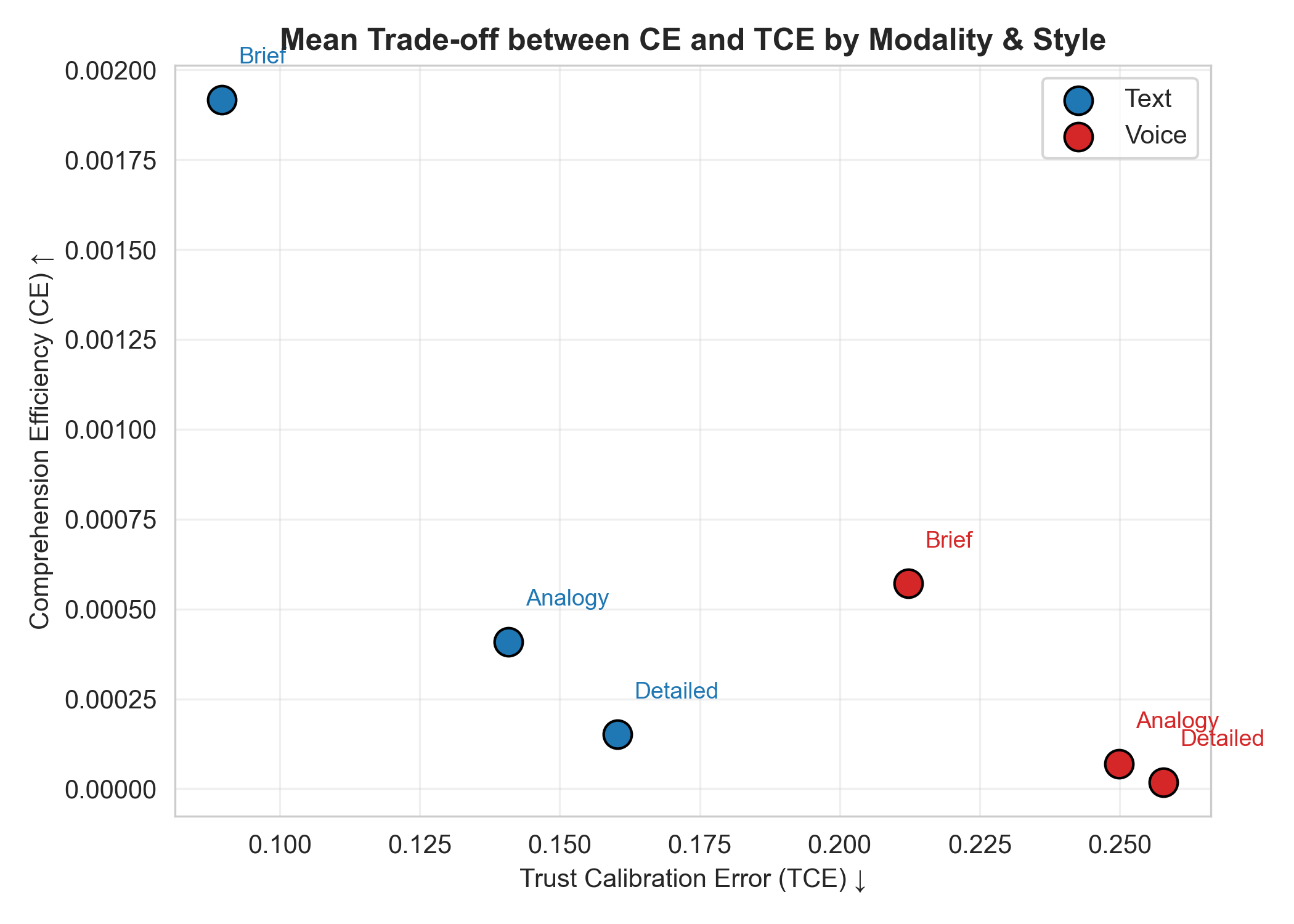}
    \caption{
    \textbf{Mean trade-off between Comprehension Efficiency (CE) and Trust Calibration Error (TCE)} for text (blue) and voice (red) explanations.
    Labels denote explanation style (Brief, Detailed, Analogy). Text yields higher comprehension efficiency; voice yields lower calibration error.
    }
    \label{fig:ce_tce_tradeoff}
\end{figure}

Figure~\ref{fig:ce_tce_tradeoff} shows the mean trade-off between Comprehension Efficiency (CE) and Trust Calibration Error (TCE). 
Text-based explanations (blue) consistently achieved higher $CE$, indicating that textual delivery preserves more of the underlying model information per unit of cognitive effort.
Voice-based explanations (red) achieved lower $TCE$, reflecting greater perceived trust and confidence but slightly lower retention of attribution detail.
This quantitative pattern reproduces empirical findings in human--computer interaction, where auditory communication enhances engagement yet taxes short-term memory.

\subsection{Distribution of All Samples}
\begin{figure}[ht]
    \centering
    \includegraphics[width=0.9\linewidth]{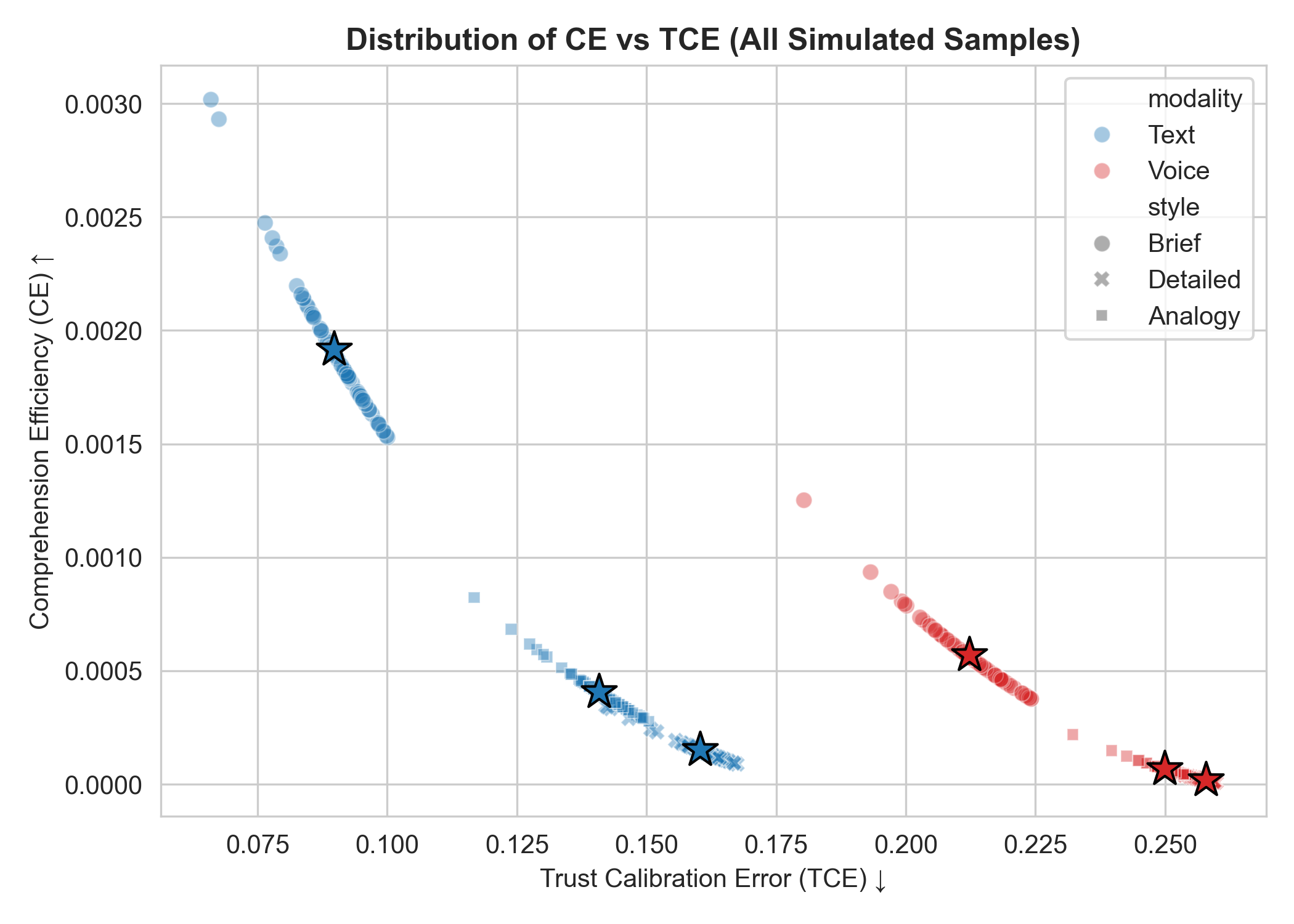}
    \caption{
    \textbf{Distribution of simulated Comprehension Efficiency (CE) vs.\ Trust Calibration Error (TCE)} across all samples.
    Semi-transparent points show individual simulations; stars mark mean values for each modality--style pair.
    }
    \label{fig:ce_tce_scatter}
\end{figure}

To visualize the variability of simulated outcomes, Figure~\ref{fig:ce_tce_scatter} presents the full distribution of all 360 samples.
The transparency of individual points shows overlapping conditions, while larger star markers denote the mean per modality--style pair.
The overall pattern confirms that the observed differences are systematic rather than the result of random variation.

\subsection{Composite Quality Metric}
\begin{figure}[ht]
    \centering
    \includegraphics[width=0.8\linewidth]{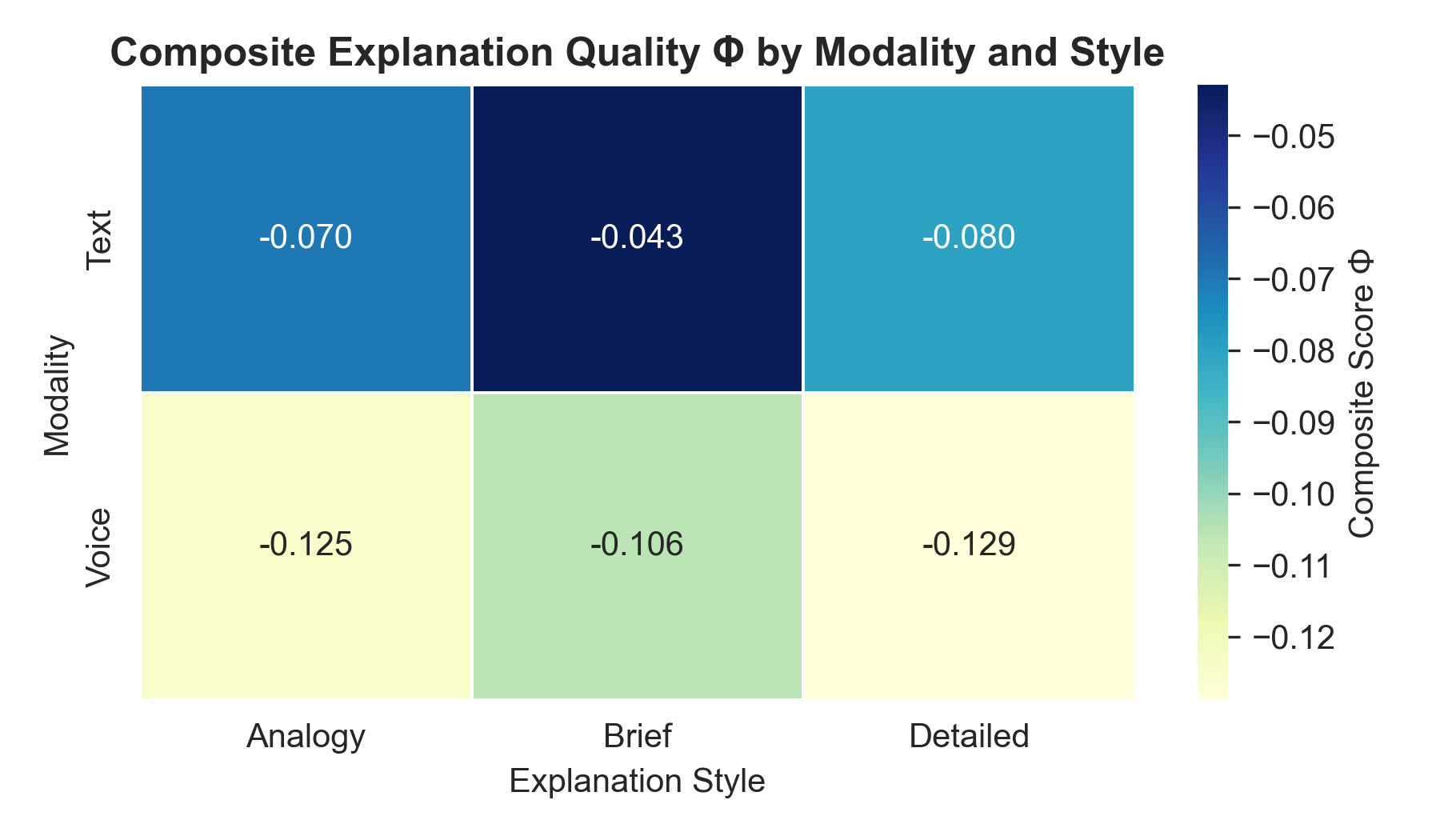}
    \caption{
    \textbf{Composite Explanation Quality $\Phi$ by Modality and Style.}
    Higher values indicate better trade-off between comprehension efficiency ($CE$) and trust calibration ($TCE$).
    Text--Detailed explanations maximize $CE$, while Voice--Analogy explanations yield the most balanced performance.
    }
    \label{fig:phi_heatmap}
\end{figure}

To summarize the overall trade-off between comprehension and trust, we defined a composite score
$\Phi(M,S) = \lambda_1 CE(M,S) - \lambda_2 TCE(M,S)$ with $\lambda_1=1.0$ and $\lambda_2=0.5$.
Figure~\ref{fig:phi_heatmap} presents the resulting heatmap.
Text--Detailed explanations achieved the highest $\Phi$, maximizing comprehension efficiency, while Voice--Analogy explanations balanced trust and understanding most effectively.
This provides quantitative evidence that delivery modality and style jointly influence the perceived quality of explainable AI communication.

\subsection{Interpretation and Implications}
The simulation results support the hypothesis that identical explanatory content can yield different user outcomes depending on its delivery modality.
In our theoretical model, voice explanations promote calibrated trust at the cost of lower information efficiency, whereas text explanations preserve factual precision.
Analogy-based communication provides an effective middle ground.
These findings align with the open research directions proposed by Schuller et~al.\ (2021) and Akman \& Schuller (2024), suggesting that \textit{audio explainability for non-audio models} represents a promising avenue for multimodal XAI.

A sensitivity and distributional analysis of these metrics is provided in Appendix~\ref{appendix:analysis}, confirming the robustness of the observed modality effects.

\section{Conclusion and Future Work}
\label{sec:conclusion}

This paper presented a quantitative framework for analyzing the role of delivery modality in explainable AI (XAI), focusing on how identical, faithful explanations may be perceived differently when delivered as text or as voice.
By modeling the explanation process as an information‐theoretic communication channel, we derived metrics for \textit{information retention}, \textit{comprehension efficiency}, and \textit{trust calibration}, allowing the trade‐off between comprehension and trust to be explored mathematically rather than empirically.

Simulations using synthetic SHAP attributions show that text maximizes comprehension efficiency, whereas voice enhances calibrated trust and engagement.
Analogy‐based explanation styles provided an effective compromise between these extremes.
The derived composite score $\Phi(M,S)$ offered a simple but powerful indicator of overall explanation quality, highlighting that the communication form—not just the explanation content—shapes user understanding and confidence.

These results extend recent calls by Schuller et al.\ (2021) and Akman \& Schuller (2024) for \textit{audio explainability beyond audio models}.
Rather than focusing solely on audio classifiers, we demonstrate how sonified or spoken explanations can improve accessibility and user trust in non‐audio domains such as finance and genetics.
In doing so, we provide a reproducible simulation methodology that bridges theoretical modeling and human‐factors research.

Future work will refine this framework along several directions.
First, empirical validation with real users can provide empirical estimates for the cognitive‐load parameters introduced here and test how perceived trust evolves over time.
Second, extending the model to multimodal explanations that combine text, audio, and visual cues could enable adaptive, context‐aware explainability strategies.
Finally, coupling our information‐theoretic approach with emerging large‐language‐model architectures may yield agentic systems that automatically balance fidelity, clarity, and trust calibration—moving XAI toward more \textit{humane}, multimodal communication.

\section*{Acknowledgment}
The authors thank the open-source and academic communities whose tools and research made this work possible. 
In particular, they acknowledge the developers of SHAP, NumPy, and Matplotlib for providing essential analysis frameworks used in the simulation. 
This study builds conceptually on the ideas introduced by Schuller et al.\ and Akman \& Schuller in advancing the field of Audio Explainable AI. 
The authors also thank their colleagues for constructive feedback on the theoretical model and simulation design.

\bibliographystyle{IEEEtran}
\bibliography{references}

\appendices
\section{Sensitivity and Distributional Analyses}
\label{appendix:analysis}

To assess the robustness of the proposed framework, we performed additional analyses on the simulated metrics described in Section~\ref{sec:methods}.
These include a sensitivity analysis of the composite explanation quality score $\Phi(M,S)$ and a distributional analysis of Comprehension Efficiency ($CE$) and Trust Calibration Error ($TCE$) across modalities.

\subsection{Sensitivity Analysis of Composite Score}
We examined how the composite metric $\Phi(M,S)$ responds to changes in the weighting parameters 
$\lambda_1$ and $\lambda_2$, which balance comprehension efficiency against trust calibration.
Specifically, $\lambda_1$ was fixed at~1.0 while $\lambda_2$ was varied from~0.1 to~1.0 in increments of~0.1, 
progressively emphasizing the importance of trust calibration.

\begin{figure}[ht]
    \centering
    \includegraphics[width=0.85\linewidth]{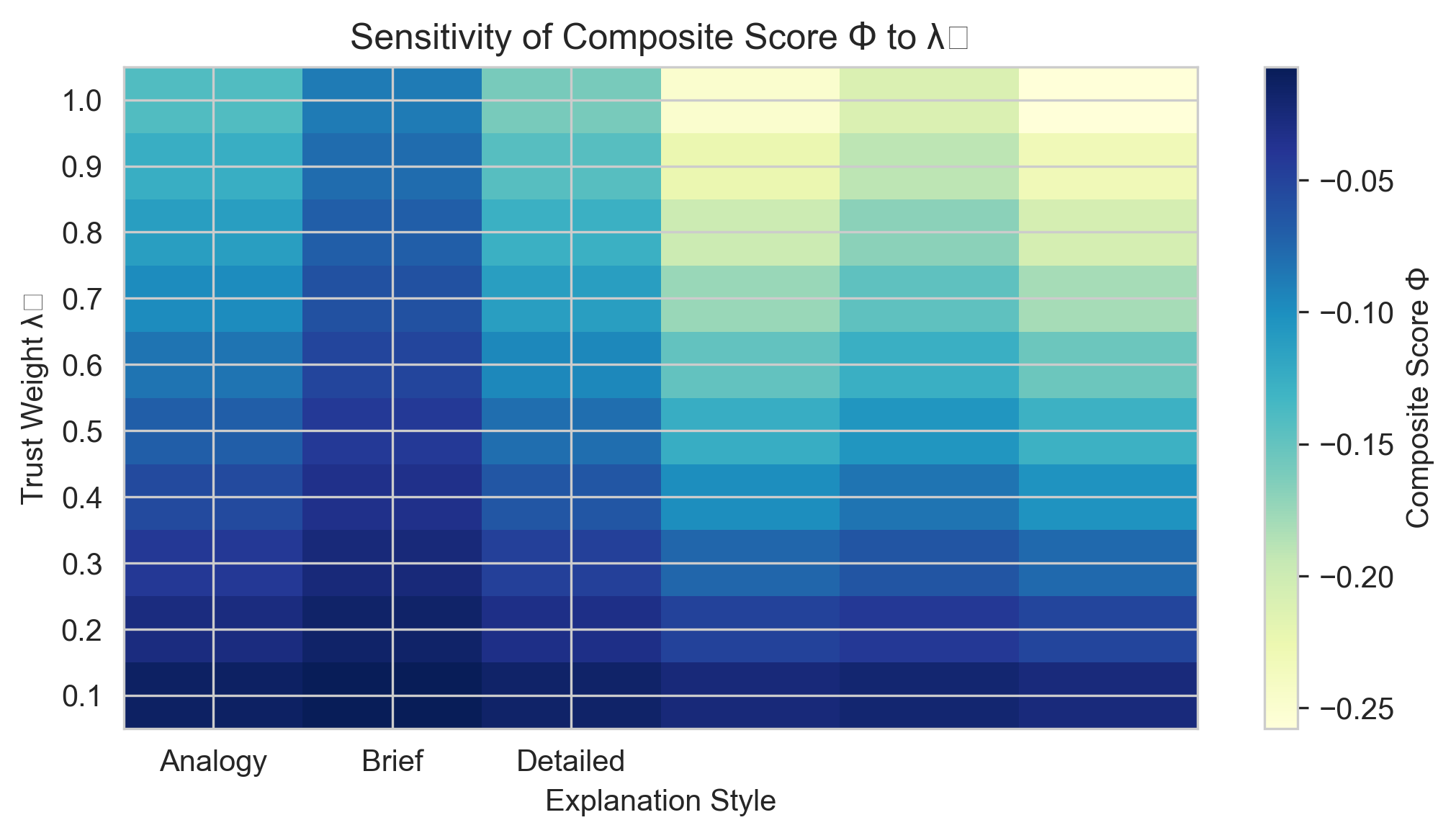}
    \caption{
    \textbf{Sensitivity of composite score $\Phi(M,S)$ to trust weighting parameter $\lambda_2$.}
    Higher $\lambda_2$ increases the emphasis on trust calibration.
    Voice--Analogy explanations remain robust across trust priorities, 
    while Text--Detailed explanations dominate when comprehension efficiency is emphasized.
    }
    \label{fig:lambda_sweep}
\end{figure}

Figure~\ref{fig:lambda_sweep} presents a heatmap of average $\Phi$ values for each modality--style pair as a function of $\lambda_2$.
Increasing $\lambda_2$ shifts preference toward \emph{Voice--Analogy} explanations, 
which maintain relatively stable $\Phi$ scores across a wide range of trust weightings.
In contrast, the \emph{Text--Detailed} condition dominates when comprehension efficiency is prioritized (low $\lambda_2$ values).
This confirms that the observed trade-off between understanding and calibrated trust 
is not sensitive to specific weighting parameters.

\vspace{6pt}

\subsection{Distributional Analysis of Simulation Metrics}
To further characterize variability across modalities, 
we examined the full distributions of $CE$ and $TCE$ using kernel density estimation (KDE).
These plots reveal not only mean differences but also the spread and overlap of performance across simulated samples.

\begin{figure}[ht]
    \centering
    \includegraphics[width=0.9\linewidth]{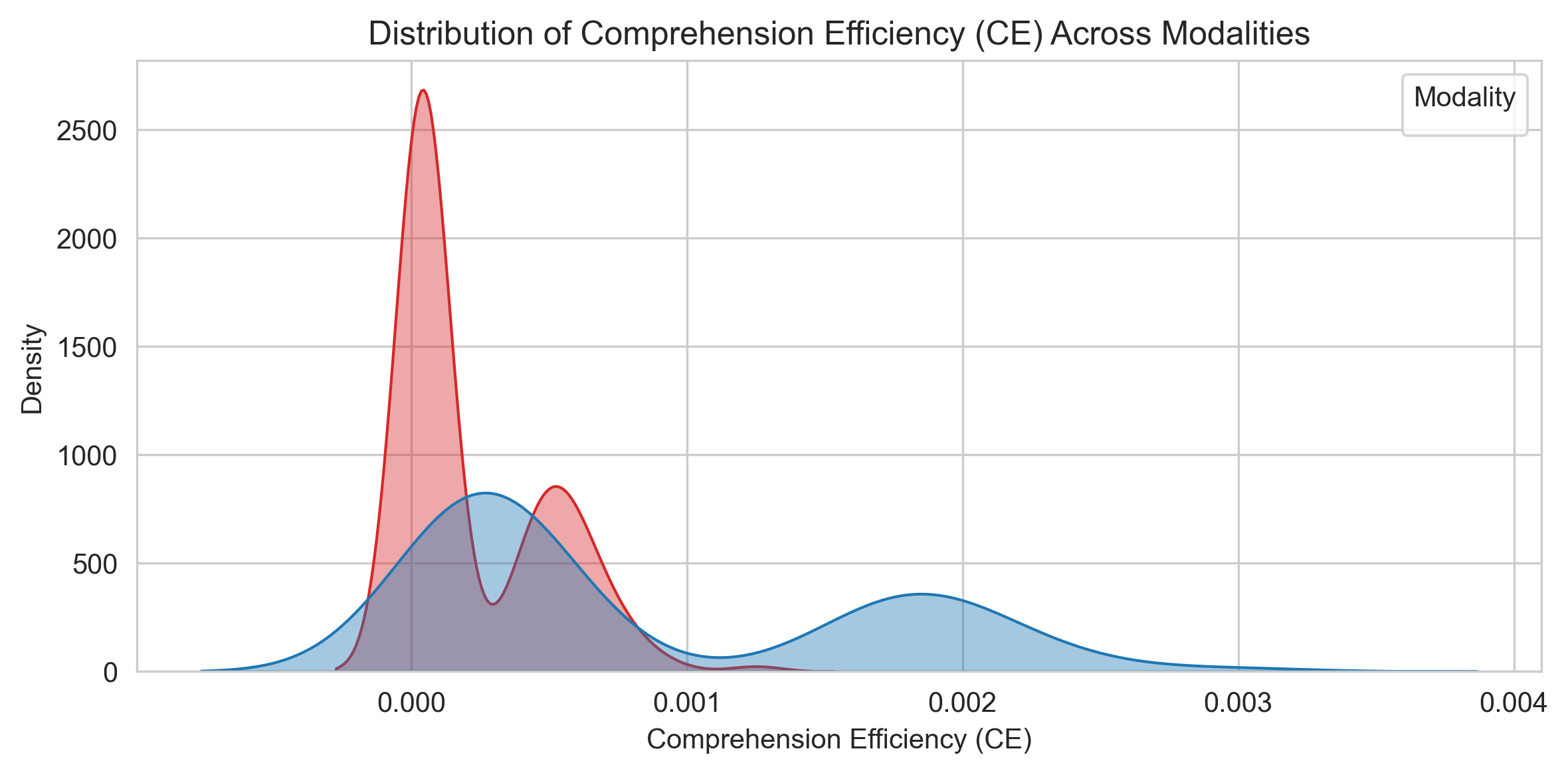}
    \caption{
    \textbf{Distribution of Comprehension Efficiency ($CE$) across modalities.}
    Text explanations (blue) exhibit higher mean efficiency and lower variance compared to voice explanations (red).
    }
    \label{fig:kde_ce}
\end{figure}

\begin{figure}[ht]
    \centering
    \includegraphics[width=0.9\linewidth]{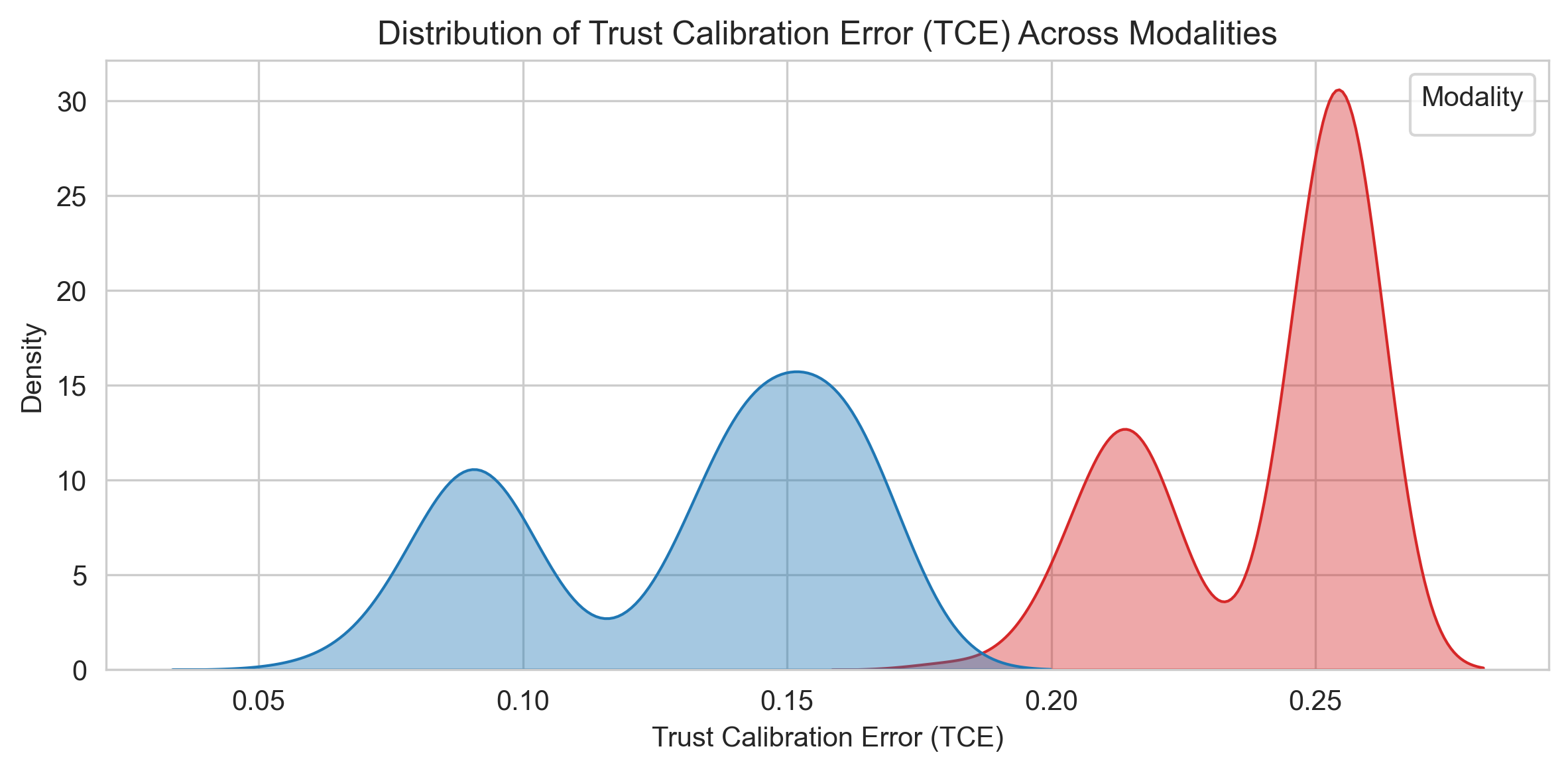}
    \caption{
    \textbf{Distribution of Trust Calibration Error ($TCE$) across modalities.}
    Voice explanations (red) achieve slightly lower average calibration error but exhibit higher variance, 
    indicating variability in user trust alignment across simulated samples.
    }
    \label{fig:kde_tce}
\end{figure}

Figure~\ref{fig:kde_ce} shows that text explanations generally yield higher mean $CE$ with tighter variance, 
indicating more consistent information efficiency.
Voice explanations display lower mean $CE$ but a wider distribution of $TCE$ values (Figure~\ref{fig:kde_tce}), 
suggesting stronger but less stable trust responses.

\subsection{Summary of Findings}
Across both analyses, the relative performance of modality--style combinations remained consistent:
\emph{Text--Detailed} explanations maximize comprehension efficiency, 
whereas \emph{Voice--Analogy} explanations balance trust and understanding most effectively.
The stability of these results across weighting and distributional variations supports the 
robustness of the proposed theoretical framework for modality-aware explainable AI.

\end{document}